# Deep Transform: Cocktail Party Source Separation via Complex Convolution in a Deep Neural Network

Andrew J.R. Simpson [#1]

[#] Centre for Vision, Speech and Signal Processing, University of Surrey
Guildford, UK
[1] Andrew.Simpson@Surrey.ac.uk

*Abstract*—Convolutional deep neural networks (DNN) are state of the art in many engineering problems but have not yet addressed the issue of how to deal with complex spectrograms. Here, we use circular statistics to provide a convenient probabilistic estimate of spectrogram phase in a complex convolutional DNN. In a typical cocktail party source separation scenario, we trained a convolutional DNN to re-synthesize the complex spectrograms of two source speech signals given a complex spectrogram of the monaural mixture – *a discriminative deep transform* (DT). We then used this complex convolutional DT to obtain probabilistic estimates of the magnitude and phase components of the source spectrograms. Our separation results are on a par with equivalent binary-mask based non-complex separation approaches.

## I. Introduction

Convolutional deep neural networks (DNN) [1]-[3] are capable of exploiting geometric assumptions about data structure in order to share network weights. When the convolutional DNN is applied in a sliding window fashion, predictions for a given datapoint may be made for multiple alternate (windowed) contexts and from this distribution of predictions a probabilistic estimate may be obtained. In computer vision problems [1]–[3], the datapoints representing pixel intensities are positive real numbers. Therefore, if the convolutional DNN is used to make predictions about intensity for a given pixel, parametric statistics may be used to obtain a probabilistic estimate of pixel intensity [4] that summarizes the predictions made in the various different contexts of the sliding window (containing the pixel in question).

While the audio spectrogram provides an intuitive visual counterpart to the 2D image of the computer vision problem, the equivalent application of convolutional DNNs to computer audition is less straight forward. In particular, the equivalence of the 2D spectrogram and the 2D image holds only for interpretations of the complex spectrogram that are limited to the magnitude component [5]-[7]. The approach of ignoring the phase component of the complex spectrogram does not imply serious limitations for classification problems (where phase may not be critical) but for audio synthesis phase is critical. Furthermore, if the convolutional DNN is applied to the phase component of the spectrogram it is not appropriate to compute probabilistic estimates of phase (from the overlapping windowed predictions) using parametric statistics. Here, we use circular statistics to obtain probabilistic estimates of phase computed using a convolutional DNN. We illustrate our approach in the context of a typical cocktail party source separation problem featuring complex spectrograms.

We trained a complex convolutional DNN to separate and re-synthesize speech in a two-speaker cocktail party scenario [7]-[8]. The DNN was used as a complex convolutional deep transform (DT) and trained to separate the speech of two concurrent speakers. At the input layer, the DT DNN was provided the complex spectrogram. At the output layer, the DT DNN was trained to re-synthesize the respective complex spectrograms of the two speakers. We then used the trained complex convolutional DT to make probabilsitic predictions about new concurrent speech mixtures of the same speakers. Using objective source separation quality metrics, we analyzed the separation quality. Our results are on a par with equivalent convolutional-DNN probabilsitic binary-mask source separation techniques [7] but offer slightly better sound quality.

## II. Method

We consider a typical simulated cocktail party listening scenario featuring a male voice and a female voice speaking concurrently. The speakers were each separately recorded (in mono) reading from a story. The two speech signals were then equalized in intensity and linearly summed (superposed) to produce a competing voice scenario.

The speech signals were decimated to a sample rate of 4 kHz and transformed into spectrograms using the short-time Fourier transform (STFT) with window size of 128 samples, overlap interval of 1 sample and a Hanning window. This provided complex spectrograms with 65 frequency bins. The spectrograms computed from the first 2 minutes of the speech signals were used as training data. A subsequent 10 seconds of data were held back for later use in testing the separation of the model.

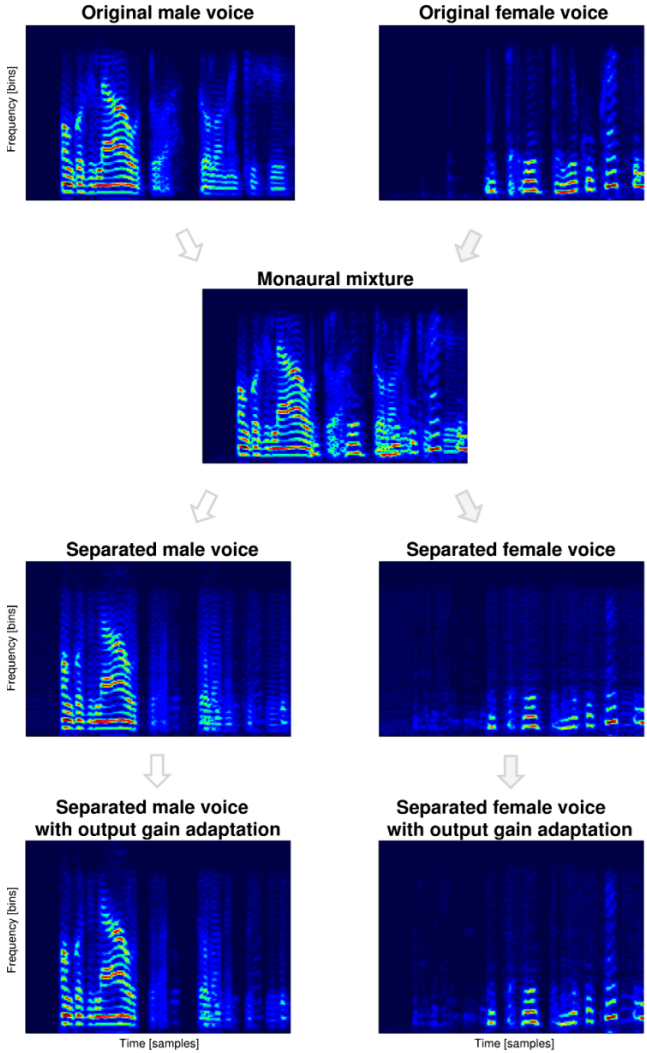

**Fig. 1. Probabilistic cocktail party source separation via complex convolutional deep transform.** The upper pair of spectrograms plot a ~2-second excerpt from the original (test) speech audio. The single (central) spectrogram plots the linear mixture of the two speech signals. The lower pairs of spectrograms plot the respective source signals separated and re-synthesized using complex CDT both with and without output gain adaptation.

For training data, the mixture and component speech spectrograms were cut up into windows (i.e., in time) of 20 samples. The windows overlaped at intervals of 10 samples. Thus, for every 20-sample window, for training the model there was mixture spectrogram matrix of size 65x20 samples and a corresponding pair of male/female source spectrograms. This gave approximately 50,000 training examples. For the testing stage, 10 seconds of speech spectrogram was used at overlap intervals of 1 sample, giving approximately 40,000 test frames (which would ultimately be applied in an overlaping convolutional output stage). Prior to windowing, the complex spectrogram data was separated into magnitude and phase spectrograms respectively, and each spectrogram was normalized to unit scale (i.e., both the magnitude and phase data were mapped to the range [0,1]).

We used a feed-forward DNN of size 2600x2600x5200 units (65 x 20 x 2 = 2600, and 2600 x 2 = 5200). Each spectrogram (magnitude and phase) window of size 65 x 20 was unpacked into a vector of length 1300, giving two respective vectors of the same size. The DNN was configured such that the input layer was a vector concatenated from the mixture magnitude spectrogram (1300 samples) followed by the mixture phase spectrogram (1300 samples), giving a total vector of length 2600 (1300 x 2). The output layer was trained to synthesize a vector featuring the sequential concatenation of the magnitude and phase spectrograms of the respective male and female speech component signals. This meant that the DNN was trained to re-synthesize the respective components into the respective concatenated locations in the output layer (vector). The DNN employed the biased-sigmoid activation function [9] throughout with zero bias for the output layer. The DNN was trained using 500 full iterations of stochastic gradient descent (SGD). Each iteration of SGD featured a full sweep of the training data. Dropout was not used in training.

*DT Probabilistic Re-Synthesis.* In the testing stage the model was used as a feed-forward signal processing device and the output layer activations were taken as synthetic output. For the test data, there was an overlap interval of 1 sample. This means that the test data described the speech spectrogram in terms of a sliding window and the output of the model was in the same sliding window format (i.e., it was a convolutional model). Each frame of input mixture spectrogram was passed through the model to produce the respective predictions of the respective magnitude and phase spectrograms for the male and female voice respectively.

An SGD trained autoencoder type DNN can feature neurons in the output layer with some degree of invariant or persistent activity [4]. To account for this activity, as in [4], we included an 'output gain adaptation' stage, where the mean activation across all test frames was subtracted from the individual activations for each frame (i.e., there was no time constant). These 'adapted' output layer predictions were then accumulated as the sliding window, $W$, (size 20) moved by 1-sample steps. Thus, 20 separate predictions were obtained for each column of the output spectrograms. For each of the male/female speakers, and for each of the respective magnitude and phase spectrograms, this gave an output distribution of spectrograms contained within a 3D matrix indexed using time ($t$), frequency ($f$) and window index ($w$).

From the 3D magnitude matrices ($M$), the running average magnitude spectrogram ($\bar{M}$) was calculated as;

$$\overline{M_{t,f}} = \frac{1}{W} \sum_{w=1}^{W} M_{t,f,w}$$
(1)

From the 3D phase spectrogram matrices ($\theta$) an equivalent circular mean phase angle was computed as follows. The phase spectrograms predicted in the output layer were first remapped from the range [0,1] to the range [0, $2\pi$] (not shown here for convenience) and then transformed into a matrix of unit vectors ($R$) in the 2D plane using the following element-wise matrix operation (subscript indices $t,f,w$ are dropped for convenience);

$$R = \begin{pmatrix} \cos(\theta \bmod 2\pi) \\ \sin(\theta \bmod 2\pi) \end{pmatrix}$$
(2)

The circular sum ($\bar{R}$), over window size ($W = 20$), is computed in an element-wise matrix operation;

$$\overline{R_{t,f}} = \sum_{w=1}^{W} R_{t,f,w}$$
(3)

Then, in a further element-wise matrix operation, the circular mean angle matrix ($\bar{\theta}$) is computed by taking the four-quadrant inverse tangent (indices from here onwards are dropped for convenience);

$$\bar{\theta} = atan2(\bar{R})$$
(4)

The respective estimated magnitude ($\bar{M}$) and phase ($\bar{\theta}$) spectrograms were then recombined into a complex spectrogram ($\hat{S}$) with the following element-wise matrix operation;

$$\hat{S} = \bar{M} \exp(\bar{\theta} i)$$
(5)

The estimated complex spectrogram ($\hat{S}$) was then subjected to an inverse STFT using an overlap-and-add procedure. Separation quality for the resulting separated audio (with respect to the original time domain audio signals) was measured using the BSS-EVAL toolbox [10] and is quantified in terms of signal-to-distortion ratio (SDR), signal-to-interference ratio (SIR) and signal-to-artefact ratio (SAR). Separation quality was assessed after each iteration of SGD training in order to evaluate the trajectory of performance (with training) in each measure.

## III. RESULTS

Fig. 1 plots spectrograms illustrating the stages of mixture and separation for a brief excerpt (~2 seconds) from the test data. The model had been trained for 500 iterations (i.e., $N = 500$). The top spectrograms plot the original male and female speech audio. The single (central) spectrogram plots the linear mixture, illustrating a large degree of overlap in this feature space. Next (downwards) are plotted the spectrograms for the complex convolutional DT probabilistic re-synthesized audio where no output gain adaptation was employed. Finally, at the bottom are plotted spectrograms representing the respective complex convolutional DT probabilistic re-synthesized audio featuring output gain adaptation.

Both sets of output spectrograms illustrate features which closely resemble those of the original signals. By visual inspection, the output gain adaptation results in less noise and somewhat better definition of the features. Both output spectrograms appear to have captured most of the bandwidth of the original signals but the fricative (noise) components of the originals have not been replicated very faithfully in either case. Informal listening revealed that the respective separated output audio was of good quality but was noisy (as Fig. 1 illustrates) in the case of the model that did not feature output gain adaptation (this model was not tested further).

Fig. 2 plots the mean objective separation quality measures (SDR, SIR, SAR), computed over the entire 10-second test data (with respect to the original audio) and averaged across the male and female voice signals, as a function of training iteration ($N$) between the range of 1 and 500. All three functions appear to be nearing convergence after around 300 iterations and all three functions are more or less monotonic. Peak separation quality (around 500 iterations), averaged across the two voices, is SDR: 8.1, SIR: 17.2, SAR: 8.9 dB. This compares well with the equivalent DNN binary mask approach reported previously (for identical data and conditions), which achieved slightly worse artefact performance (for an equivalent SIR of 17.2 – see Fig. 2 of [7]); SDR: 7.5, SAR: 8.2 dB. This minor advantage is presumably due to the ability of the present approach to employ phase directly, where in the convolutional DNN binary mask approach phase is taken from the mixture spectrogram. The present results also compare reasonably well with the ideal binary mask computed for the same test data (i.e., using a mask computed from the source spectrograms), which achieves SDR: 11.3, SIR: 21.6, SAR: 11.7. The present results are also much better than the time-domain convolutional DT probabilistic re-synthesis approach reported previously [11]. To some degree, it may be that the inclusion of both phase and magnitude in the present approach provides some small advantage in terms of sampling that may be broadly equivalent to over-sampling [12].

We also note that, similar to the binary-mask based convolutional DNN approach reported previously [7], the present performance appears superior to earlier non-negative matrix factorization (NMF) based approaches [5], [6], which featured small scale DNN within the NMF pipeline. However, the results of the NMF-based methods are not directly comparable with the present results.

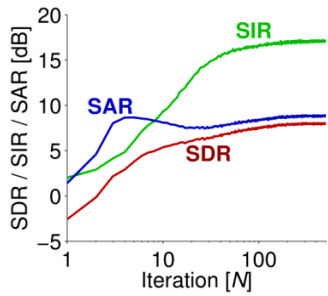

**Fig. 2. Complex convolutional DT: Separation quality as a function of training iterations (*N*).** Mean signal-to-interference ratio (SIR), signal-to-distortion ratio (SDR) and signal-to-artefact ratio (SAR) computed from the audio separated using complex convolutional DT as a function of training iteration (*N*). The measures were computed from the 10-second test audio using the BSS-EVAL toolkit [10] and averaged across the two voices.

## IV. Discussion and Conclusion

We have introduced a complex convolutional DT approach to cocktail party source separation using spectrograms. Our probabilistic approach features both parametric statistical estimation of spectrogram magnitude and circular statistical estimation of phase. The convolutional DNN was trained on two minutes of the speech of two speakers and tested on 10 seconds of new speech from the same speakers. Separation quality is similar to binary mask based convolutional DNN aproaches but features slightly improved artefact performance.

Although the DNN employed here is of only 3 layers, if we consider the degree of abstraction already provided by the STFT and inverse STFT (giving an effective 'depth' of 5 layers of demodulation [9] and synthesis [4]) then it is not surprising that the approach works as well as it does [7]. Furthermore, the ability of the model to operate with full phase information, whilst retaining the 2D topographic projection of the STFT, appears an advantage. It may also be that performance in the present model is further enhanced by the use of the Hanning window, which acts (similarly to over-sampling) to mitigate aliasing that has been suggested to affect DNN learning and performance [12]. More generally, the circular statistical process of probabilistic synthesis described here may be useful for more general probabilistic synthesis at the level of the complex spectrogram.

At a more general level, if our model is interpreted as an auditory model, the featured output gain adaptation appears similar to that observed in the early auditory system [13]-[18]. Auditory gain adaptation is temporal and occurs on various timescales up to tens of seconds and even minutes [13]-[18]. In principle, the output gain adaptation of the present model may be interpreted as featuring a rectangular temporal integration window of length 10 seconds (the entire test data). Hence, the present model may be interpreted as demonstrating that neuronal output gain adaptation may be useful in terms of synthesis noise reduction. Hence, our findings may provide some insight into the neural processing of the auditory system and its associated perceptual illusions and function.

ACKNOWLEDGMENT

AJRS was supported by grant EP/L027119/1 from the UK Engineering and Physical Sciences Research Council (EPSRC).